\documentclass[12pt,a4paper]{article}
\usepackage{amssymb,latexsym,amsmath,amsthm}
\usepackage[dvips]{color}
\usepackage{graphicx}
\usepackage[mathscr]{eucal}
\begin{document}
\baselineskip=0.3in
\def\appendix{\par \setcounter{section}{0} \def\thesection{APPENDIX:}}
\newdimen\jot \jot=8mm

\title{Nonparametric Methodology for the Time-Dependent Partial Area under the ROC Curve
}
\author{
  Hung Hung  \hspace{1cm} Chin-Tsang Chiang\\
  National Taiwan University}
\date{\empty}
\maketitle

\begin{abstract}

To assess the classification accuracy of a continuous diagnostic
result, the receiver operating characteristic (ROC) curve is
commonly used in applications. The partial area under the ROC curve
(pAUC) is one of widely accepted summary measures due to its
generality and ease of probability interpretation. In the field of
life science, a direct extension of the pAUC into the time-to-event
setting can be used to measure the usefulness of a biomarker for
disease detection over time. Without using a trapezoidal rule, we
propose nonparametric estimators, which are easily computed and have
closed-form expressions, for the time-dependent pAUC. The asymptotic
Gaussian processes of the estimators are established and the
estimated variance-covariance functions are provided, which are
essential in the construction of confidence intervals. The finite
sample performance of the proposed inference procedures are
investigated through a series of simulations. Our method is further
applied to evaluate the classification ability of CD4 cell counts on
patient's survival time in the AIDS Clinical Trials Group (ACTG) 175
study. In addition, the inferences can be generalized to compare the
time-dependent pAUCs between patients received the prior
antiretroviral therapy and those without it.

\end{abstract}

\section{Introduction}

\renewcommand{\theequation}{1.\arabic{equation}}
\setcounter{equation}{0}

Decision-making is an important issue in many fields such as signal
detection, psychology, radiology, and medicine. For example,
preoperative diagnostic tests are medically necessary and
implemented in clinical preventive medicine to determine those
patients for whom surgery is beneficial. For the sake of cost-saving
or performance improvement, new diagnostic tests are often
introduced and the classification accuracies of them are evaluated
and compared with the existing ones. The ROC curve, a plot of the
true positive rate (TPR) versus the false positive rate (FPR) for
each possible cut point, has been widely used for this purpose when
the considered diagnostic tests are continuous. One advantage of the
ROC curve is that it describes the inherent classification
capability of a biomarker without specifying a specific threshold.
Moreover, the invariance characteristic of ROC curve in measurement
scale provides a suitable base to compare different biomarkers.
Generally, the more the curve moves toward the point $(0,1)$, the
better a biomarker performs.

In many applications, the area under the ROC curve (AUC), one of the
most popular summary measures of the ROC curve, is used to evaluate
the classification ability of a biomarker. It has the probability
meaning that the considered biomarker of a randomly selected
diseased case is greater than that of a non-diseased one. Generally,
a perfect biomarker will have the AUC of one while a poor one takes
a value close to 0.5. Since the AUC is the whole area under the ROC
curve, relevant information might not be entirely captured in some
cases. For example, two crossed ROC curves can have the same AUC but
totally different performances. Furthermore, there might be limited
or no data in the region of high FPR. In view of these drawbacks, it
is more useful to see the pAUC within a certain range of TPR or FPR.
To evaluate the performance of several biomarkers, McClish (1989)
adopted the summary measure pAUC for the FPR over a practically
relevant interval. On the other hand, Jian, Metz, Nishikawa (1996)
showed that women with false-negative findings at mammography cannot
be benefited from timely treatment of the cancer. Thus, these
authors suggested using the pAUC with a portion of the true positive
range in their applied data. Although their perspectives are
different, the main frame is the same: only the practically
acceptable area under the ROC curve is assessed. As mentioned by
Dwyer (1997), the pAUC is a regional analysis of the ROC curve
intermediate between the AUC and individual points on the ROC curve.
The pAUC becomes a good measure of classification accuracy because
it is easier for a practitioner to determine a range of TPR or/and
FPR that are relevant. Several estimation and inference procedures
have been proposed by Emir, Wieand, Jung, and Ying (2000), Zhang,
Zhou, Freeman, and Freeman (2002), Dodd and Pepe (2003), among
others. A more thorough understanding of the ROC, AUC, and pAUC can
be also found in Zhou, McClish, and Obuchowski (2002).

Recent research in ROC methodology has extended the binary disease
status to the time-dependent setting. Let $T$ denote the time to a
specific disease or death and $Y$ represent the continuous
diagnostic marker measured before or onset of the study with joint
survivor function $S(t,y)=P(T>t, Y>y)$. For each fixed time point
$t$, the disease status can be defined as a case if $T\leq t$ and a
control otherwise. To evaluate the ability of $Y$ in classifying
subjects who is diseased before time $t$ or not, Heagerty, Lumley,
and Pepe (2000) generalized the traditional TPR and FPR to the
time-dependent TPR and FPR as $TPR_t(y)=P(Y>y|T\leq t)$ and
$FPR_t(y)=P(Y>y|T>t)$, which can be further derived to be
$(S(0,y)-S(t,y))/(1-S_T(t))$ and $S(t,y)/S_T(t)$ with
$S_T(t)=S(t,-\infty)$. For the time-dependent AUC, Chambless and
Diao (2006), Chiang, Wang, and Hung (2008), and Chiang and Hung
(2008) proposed different nonparametric estimators and developed the
corresponding inference procedures. As for the time-dependent pAUC,
there is still far too little research on this topic. We propose
nonparametric estimators, which are shown to converge weakly to
Gaussian processes, and the estimators for the corresponding
variance-covariance functions. The established properties facilitate
us to make inference on the time-dependent pAUC and can be
reasonably applied to the time-dependent AUC because it is a special
case of this summary measure.

The rest of this paper is organized as follows. In Section 2, the
nonparametric estimation and inference procedures are proposed for
the time-dependent pAUC. The finite sample properties of the
estimators and the performance of the constructed confidence bands
are studied through Monte Carlo simulations in Section 3. Section 4
presents an application of our method to the ACTG 175 study. In this
section, an extended inference procedure is further provided for the
comparison of the time-dependent pAUCs. Some conclusions and future
works are addressed in Section 5. Finally, the proof of main results
is followed in the Appendix.

\section{Estimation and Inferences}

\renewcommand{\theequation}{2.\arabic{equation}}
\setcounter{equation}{0}

In this section, we estimate the time-dependent pAUC and develop the
corresponding inference procedures. Without loss of generality, the
time-dependent pAUC is discussed for restricted $FPR_t(y)$ because
that for restricted $TPR_t(y)$ can be derived in the same way by
reversing the roles of case and control subjects.

\subsection{Estimation}

Let $X$ be the minimum of $T$ and censoring time $C$,
$\delta=I(X=T)$ represent the censoring status, and $q_{\alpha
t}=FPR_t^{-1}(\alpha)=\inf\{y:FPR_t(y)\leq \alpha\}$, $\alpha \in
(0,1]$, denote the $(1-\alpha)th$ quantile of $Y$ conditioning on
$\{T> t\}$ at the fixed time point $t$. Following the expression
$\{-\int TPF_t(y)d_yFPR_t(y)\}$ for the time-dependent AUC, the
time-dependent pAUC $\theta_t(q_{\alpha t})$ with the $FPR_t(y)$
less than $\alpha$ is derived to be functional of $S(t,y)$:
\begin{eqnarray}
\Theta_{\alpha}(S)=\frac{-\int(S(0,u)-S(t,u))I(u\geq q_{\alpha
t})d_uS(t,u)}{S_T(t)(1-S_T(t))}. \label{1.1}
\end{eqnarray}
Note that the value of $\theta_t(q_{\alpha t})$ for a perfect
biomarker should be $\alpha$ while a useless one is $0.5\alpha^2$.
Same with the interpretation of Cai and Dodd (2008), the rescaled
time-dependent pAUC $\theta_t(q_{\alpha t})/\alpha$ can be explained
as the probability that the test result of a case $\{T_{i}\leq t\}$
is higher than that of a control $\{T_{j}>t\}$ with its value
exceeding $q_{\alpha t}$ for $i\neq j$, i.e.,
$P(Y_{i}>Y_{j}|T_{i}\leq t, T_{j}>t, Y_{j}>q_{\alpha t})$.

From the formulation in (\ref{1.1}), an estimator of
$\theta_t(q_{\alpha t})$ can be obtained if $S(t,y)$ is estimable.
Under marker dependent censoring ($T$ and $C$ are independent
conditioning on $Y$), Akritas (1994) suggested estimating $S(t,y)$
by $\widehat{S}(t,y)=n^{-1}\sum_{i=1}^n
\widehat{S}_{T}(t|Y_i)I(Y_i>y)$, where
\begin{eqnarray}
\widehat{S}_{T}(t|y)= \prod_{\{i:X_i\leq
t,\delta_i=1\}}\{1-\frac{K_{\lambda}(\widehat{S}_Y(Y_{i})-\widehat{S}_Y(y))}{n\widehat{S}_X(X_i|y)}\}
\label{1.2}
\end{eqnarray}
is an estimator of $S_{T}(t|y)=P(T>t|Y=y)$ with
$\widehat{S}_{Y}(y)=n^{-1}\sum_{j=1}^nI(Y_j>y)$ and
$\widehat{S}_X(t|y)=n^{-1}\sum^{n}_{j=1}I(X_{j}\geq
t)K_{\lambda}(\widehat{S}_Y(Y_{j})-\widehat{S}_Y(y))$ being
estimators of $S_Y(y)=P(Y>y)$ and $S_X(t|y)=P(X>t|Y=y)$. Here,
$K_{\lambda}(u)=(2\lambda)^{-1}I(|u|<\lambda)$ and $\lambda$ is a
nonnegative smoothing parameter. Substituting $\widehat{S}(t,y)$ for
$S(t,y)$ in (\ref{1.1}), $\theta_t(q_{\alpha t})$ is proposed to be
estimated by
\begin{eqnarray}
\widehat{\theta}_t(\widehat{q}_{\alpha
t})\stackrel{\triangle}{=}\Theta_{\alpha}(\widehat{S})=\frac{n^{-2}\sum_{i\neq
j}(1-\widehat{S}_T(t|Y_i))\widehat{S}_T(t|Y_j)\phi_{ij}(\widehat{q}_{\alpha
t})} {\widehat{S}_T(t)(1-\widehat{S}_T(t))},\label{2.1}
\end{eqnarray}
where $\phi_{ij}(y)=I(Y_i>Y_j>y)$, $\widehat{q}_{\alpha
t}=\widehat{FPR}_t^{-1}(\alpha)$,
$\widehat{FPR}_t(y)=\widehat{S}(t,y)/\widehat{S}_T(t)$, and
$\widehat{S}_T(t)=\widehat{S}(t,-\infty)$. In the Appendix, we show
that $\sqrt{n}(\widehat{\theta}_t(\widehat{q}_{\alpha
t})-\theta_t(q_{\alpha t}))$ is uniformly approximated by
$n^{-1/2}\sum_{i=1}^n\Psi_{\alpha i}(t)$ and converges weakly to a
mean zero Gaussian process with variance-covariance function
$\Sigma_{\alpha}(s,t)=E[\Psi_{\alpha i}(s)\Psi_{\alpha i}(t)]$ for
$t\in(0,\tau]$ and $P(X>\tau)>0$. The application of kernel function
$K_{\lambda}(u)$ provides the nearest neighbor estimator of
$S(t,y)$. An alternative choice of kernel function is possible and
will lead to a different estimator of $\theta_t(q_{\alpha t})$. As
mentioned in Akritas (1994), the asymptotic properties of
$\widehat{S}(t,y)$ is irrelevant to the choice of kernel function
under some regularity conditions and so is
$\widehat{\theta}_t(\widehat{q}_{\alpha t})$. The author further
showed that any other estimator for $S(t,y)$ is at least as
dispersed as $\widehat{S}(t,y)$ and the choice of $\lambda$ is
irrelevant to the measurement scale of $Y$. It is not difficult to
see that the estimation problem of $\theta_t(q_{\alpha t})$ becomes
that of $S(t,y)$. From this perspective, the proposed estimation
procedure can be extended to any censoring or truncation mechanisms
provided that $S(t,y)$ is estimable.

Alternatively, one might be interested in making inference on the
time-dependent pAUC $(\theta_t(q_{\alpha' t})-\theta_t(q_{\alpha
t}))$ over the range $[\alpha,\alpha']$, $0\leq \alpha<\alpha'\leq
1$, of $FPR_{t}(y)$. The estimator
$(\widehat{\theta}_t(\widehat{q}_{\alpha'
t})-\widehat{\theta}_t(\widehat{q}_{\alpha t}))$ is suggested and
the limiting Gaussian process of
$\sqrt{n}\{(\widehat{\theta}_t(\widehat{q}_{\alpha'
t})-\widehat{\theta}_t(\widehat{q}_{\alpha t}))-
(\theta_t(q_{\alpha' t})-\theta_t(q_{\alpha t}))\}$ is a direct
consequence of the large sample property of
$\widehat{\theta}_t(\widehat{q}_{\alpha t})$. When the complete
failure time data $\{T_i,Y_i\}_{i=1}^n$ are available, $S(t,y)$ can
be estimated by an empirical estimator
$\widetilde{S}(t,y)=n^{-1}\sum_{i=1}^nI(T_i>t,Y_i>y)$. A natural
estimator for $\theta_t(q_{\alpha t})$ is obtained as
\begin{eqnarray}
\widetilde{\theta}_t(\widetilde{q}_{\alpha
t})=\frac{n^{-2}\sum_{i\neq j}I(T_i\leq
t,T_j>t)\phi_{ij}(\widetilde{q}_{\alpha t})}
{\widetilde{S}_T(t)(1-\widetilde{S}_T(t))},\label{2.7}
\end{eqnarray}
where $\widetilde{q}_{\alpha t}=\widetilde{FPR}_t^{-1}(\alpha)$,
$\widetilde{FPR}_t(y)=\widetilde{S}(t,y)/\widetilde{S}_T(t)$, and
$\widetilde{S}_T(t)=\widetilde{S}(t,-\infty)$. By substituting the
disease and disease-free groups for the time-varying case and
control ones, $\theta_{t}(q_{\alpha t})$ and the estimator
$\widetilde{\theta}_t(\widetilde{q}_{\alpha t})$ will reduce to the
time-invariant pAUC and the nonparametric estimator of Dodd and Pepe
(2003). By the similar argument as in the proof of the asymptotic
Gaussian process of $\widehat{\theta}_t(\widehat{q}_{\alpha t})$, it
is straightforward to derive that
$\sqrt{n}(\widetilde{\theta}_t(\widetilde{q}_{\alpha
t})-\theta_t(q_{\alpha t}))$ converges weekly to a Gaussian process
with mean zero and variance-covariance function
$\Sigma^{*}_{\alpha}(s,t)=E[\Psi^{*}_{\alpha i}(s)\Psi^{*}_{\alpha
i}(t)]$, where
\begin{equation}
\Psi_{\alpha i}^*(t)=\frac{U_{i}^*(t,q_{\alpha t})+\eta(t,q_{\alpha
t})V_{i}^*(t,-\infty)+ (\alpha S_T(t)-S_Y(q_{\alpha
t}))(V_{i}^*(t,q_{\alpha t})-\alpha
V_{i}^*(t,-\infty))}{S_T(t)(1-S_T(t))},\nonumber
\end{equation}
$U_{i}^*(t,y)=E[h_{ij}^*(t,y)+h_{ji}^*(t,y)|T_i,Y_i]-2H(t,y)$,
$h_{ij}^*(t,y)=I(T_i\leq t, T_j>t)\phi_{ij}(y)$, and
$V_{i}^*(t,y)=I(T_i>t,Y_i>y)$.

\subsection{Inference Procedures on the Time-Dependent pAUC}

The confidence intervals for $\theta_t(q_{\alpha t})$ and
$(\theta_t(q_{\alpha' t})-\theta_t(q_{\alpha t}))$ can be
constructed by the asymptotic Gaussian processes and the estimated
variance-covariance functions. Replacing the parameters with their
sample analogues, $\Psi_{\alpha i}(t)$ is proposed to be estimated
by
\begin{eqnarray}
\widehat{\Psi}_{\alpha
i}(t)=\frac{\widehat{U}_{i}(t,\widehat{q}_{\alpha t})+
\widehat{\eta}(t,\widehat{q}_{\alpha t})\widehat{V}_{i}(t,-\infty)+
(\alpha\widehat{S}_T(t)-\widehat{S}_Y(\widehat{q}_{\alpha
t}))(\widehat{V}_{i}(t,\widehat{q}_{\alpha t})
-\alpha\widehat{V}_{i}(t,-\infty))}{\widehat{S}_T(t)(1-\widehat{S}_T(t))},
\label{2.2}
\end{eqnarray}
where $\widehat{U}_{i}(t,y)=n^{-1}\sum_{\{j:j\neq
i\}}(\widehat{h}_{ij}(t,y)+\widehat{h}_{ji}(t,y))-2\widehat{H}(t,y)
+(\widehat{S}_Y(Y_i)-\widehat{S}(t,y))\widehat{\xi}_i(t)I(Y_i>y)$,
$\widehat{V}_{i}(t,y)=(\widehat{S}_T(t|Y_i)+\widehat{\xi}_i(t))I(Y_i>y)-\widehat{S}(t,y)$,
and
$\widehat{\eta}(t,y)=\widehat{H}(t,y)(2\widehat{S}_T(t)-1)/(\widehat{S}_T(t)-\widehat{S}_T^2(t))$
with
$\widehat{h}_{ij}(t,y)=(1-\widehat{S}_T(t|Y_i))\widehat{S}_T(t|Y_j)\phi_{ij}(y),
\widehat{H}(t,y)=n^{-2}\sum_{i\neq j}\widehat{h}_{ij}(t,y)$,
$\widehat{\xi}_i(t)=-\widehat{S}_T(t|Y_i)\int_0^t\widehat{S}_{X}^{-1}
(u|Y_i)d_u\widehat{M}_i(u|Y_i)$, and $\widehat{M}_i(t|Y_i)=I(X_i
\leq t)\delta_i+\ln\widehat{S}_{T}(t\wedge X_i|Y_i)$. Thus, it is
straightforward to have an estimated variance-covariance function
\begin{eqnarray}
\widehat{\Sigma}_{\alpha}(s,t)=\frac{1}{n}\sum_{i=1}^n\widehat{\Psi}_{\alpha
i}(s)\widehat{\Psi}_{\alpha i}(t)\label{2.3}
\end{eqnarray}
and a $(1-\varsigma)$, $0<\varsigma<1$, pointwise confidence
interval for $\theta_t(q_{\alpha t})$:
\begin{eqnarray}
\widehat{\theta}_t(\widehat{q}_{\alpha
t})\pm\frac{Z_{\varsigma/2}}{\sqrt{n}}\widehat{\Sigma}_{\alpha}^{1/2}(t,t),
\label{2.4}
\end{eqnarray}
where $Z_{\varsigma/2}$ is the $(1-\varsigma/2)$ quantile value of
the standard normal distribution. With the independent and
identically distributed representation
$n^{-1/2}\sum_{i=1}^n\Psi_{\alpha i}(t)$, the re-sampling technique
of Lin, Wei, Yang, and Ying (2000) is applied to determine a
critical point $L_{\varsigma}$ so that
\begin{eqnarray}
P(\sup_{t\in[\tau_1,\tau_2]}|\frac{\sqrt{n}(\widehat{\theta}_t(\widehat{q}_{\alpha
t})-\theta_t(q_{\alpha
t}))}{\widehat{\Sigma}_{\alpha}^{1/2}(t,t)}|<L_{\varsigma})\doteq
1-\varsigma \label{2.5}
\end{eqnarray}
for a subinterval $[\tau_1,\tau_2]$ of interest within the time
period $[0,\tau]$. The validity of (\ref{2.5}) enables us to
construct a $(1-\varsigma)$ simultaneous confidence band for
$\{\theta_t(q_{\alpha t}): t\in [\tau_1,\tau_2]\}$ via
\begin{eqnarray}
\{\widehat{\theta}_t(\widehat{q}_{\alpha
t})\pm\frac{L_{\varsigma}}{\sqrt{n}}\widehat{\Sigma}_{\alpha}^{1/2}(t,t):t\in[\tau_1,\tau_2]\}.\label{2.6}
\end{eqnarray}

Note that both pointwise and simultaneous confidence bands for
$(\theta_t(q_{\alpha' t})-\theta_t(q_{\alpha t}))$ can be
constructed as the above ones. When
$\widetilde{\theta}_t(\widetilde{q}_{\alpha t})$ is applicable, the
confidence bands are easily obtained by substituting
$\widetilde{\Sigma}^{*}_{\alpha}(s,t)=n^{-1}\sum^{n}_{i=1}\widetilde{\Psi}^{*}_{\alpha
i}(s)\widetilde{\Psi}^{*}_{\alpha i}(t)$ for
$\widehat{\Sigma}_{\alpha}(s,t)$ in (\ref{2.4}) and (\ref{2.6}),
where
\begin{equation*}
\widetilde{\Psi}_{\alpha
i}^*(t)=\frac{\widetilde{U}_{i}^*(t,\widetilde{q}_{\alpha t})+
\widetilde{\eta}(t,\widetilde{q}_{\alpha t})V_{i}^*(t,-\infty)+
(\alpha\widetilde{S}_T(t)-\widehat{S}_Y(\widetilde{q}_{\alpha
t}))(V_{i}^*(t,\widetilde{q}_{\alpha t}) -\alpha
V_{i}^*(t,-\infty))}{\widetilde{S}_T(t)(1-\widetilde{S}_T(t))},
\label{2.8}
\end{equation*}
$\widetilde{U}_{i}^*(t,y)=n^{-1}\sum_{\{j:j\neq
i\}}(h_{ij}^*(t,y)+h_{ji}^*(t,y))-2\widetilde{H}(t,y)$,
$\widetilde{H}(t,y)=n^{-2}\sum_{i\neq j}h_{ij}^*(t,y)$, and
$\widetilde{\eta}(t,y)=\widetilde{H}(t,y)(2\widetilde{S}_T(t)-1)/(\widetilde{S}_T(t)-\widetilde{S}_T^2(t))$.

\section{Numerical Studies}
\renewcommand{\theequation}{3.\arabic{equation}}
\setcounter{equation}{0}

In this section, Monte Carlo simulations are conducted to
investigate the finite sample properties of the proposed estimators
and the performance of the inference procedures. The continuous
biomarker $Y$ is designed to follow a standard normal distribution.
Conditioning on $Y=y$, the failure time $T$ and the censoring time
$C$ are independently generated from a lognormal distribution with
parameters $\mu=-0.15y+\ln 10$ and $\sigma =0.3$, and an exponential
distribution with scale parameter $10b\{2I(y<0)+I(y \geq 0)\}$,
where the constant $b$ is set to produce the censoring rates of
$0\%$, $30\%$ and $50\%$. In our numerical studies, 500 data sets of
500 and 1000 observations are simulated. The estimators and the
pointwise confidence intervals of $\theta_{t}(q_{\alpha t})$ are
evaluated at the selected time points $t_{0.4}$, $t_{0.5}$, and
$t_{0.6}$ with $\alpha=$0.1, 0.2, and 0.3, where $t_p$ is the $p$th
quantile of the distribution of $T$. Moreover, the simultaneous
confidence bands for $\theta_{t}(q_{\alpha t})$ over the
subintervals $[t_{0.4},t_{0.5}]$ and $[t_{0.4},t_{0.6}]$ are
considered. Since a small portion of cases or controls occur outside
$[t_{0.4},t_{0.6}]$ under the above design, the simulation results
are presented within this time period.

When survival times are subject to censoring, an appropriate
smoothing parameter urgently becomes necessary in the estimation of
$\theta_{t}(q_{\alpha t})$. It usually attempts to select a
bandwidth that minimizes the asymptotic mean squared error of an
estimator, which is obtained by using the plug-in method for unknown
parameters. This approach, however, would lead to further bandwidth
selection problems and is infeasible in our current setting. For the
bandwidth selection, we propose a simple and easily implemented
data-driven method. This procedure is to find a bandwidth, say,
$\lambda_{opt}$ which minimizes the following integrated squared
error
\begin{eqnarray}
ISE(\lambda)=\int_0^1(\widehat{S}_e(u)-(1-u))^2dN_{ei}(u),
\label{3.1}
\end{eqnarray}
where $\widehat{S}_e(u)$ is the Kaplan-Meier estimator computed
based on the data $\{e_i,\delta_i\}_{i=1}^n$,
$e_i=1-\widehat{S}_T^{(-i)}(X_i|Y_i)$, $\widehat{S}_T^{(-i)}(t|y)$
is computed as $\widehat{S}_T(t|y)$ with the $i$th observation
$(X_{i},\delta_{i},Y_{i})$ deleted, and
$N_{ei}(u)=\delta_{i}I(e_{i}\leq u)$. The rationale behind
(\ref{3.1}) is that $\{1-S_T(X_i|Y_i),\delta_i\}_{i=1}^n$ can be
shown to be an independent censored sample from a uniform
distribution $U(0,1)$ under the validity of conditionally
independent censoring. For each generated sample,
$\widehat{\theta}_{t}(\widehat{q}_{\alpha t})$ and
$\widehat{\Psi}_{\alpha i}(t)$'s are computed by using
$\lambda_{opt}$ and the subjective bandwidths of 0.01 and 0.2. Among
the 500 simulated samples, the bandwidths obtained from minimizing
$ISE(\lambda)$ in (\ref{3.1}) have a range between 0.01 and 0.2, and
medians of about 0.09, 0.1, 0.1, 0.07, 0.08, and 0.08 for $(n,c.r.)$
of $(500,0\%)$, $(500,30\%)$, $(500,50\%)$, $(1000,0\%)$,
$(1000,30\%)$, and $(1000,50\%)$, where $n$ and $c.r.$ represent the
sample size and the censoring rate.

Tables 1-3 summarize the averages and standard deviations of
estimates, the standard errors, and the empirical coverage
probabilities of 0.95 pointwise confidence intervals for
$\theta_{t}(q_{\alpha t})$. For the complete failure time data
(i.e., $c.r.=0\%$), $\widetilde{\theta}_{t}(\widetilde{q}_{\alpha
t})$ and $\widehat{\theta}_{t}(\widehat{q}_{\alpha t})$, computed
using $\lambda_{opt}$, give separately a slight overestimate and
underestimate of $\theta_{t}(q_{\alpha t})$. Moreover, the variance
$\Sigma^{*}_{\alpha}(t,t)$ tends to be underestimated, which leads
to a lower coverage probability. As expected, the bias and standard
deviation of $\widehat{\theta}_{t}(\widehat{q}_{\alpha t})$ will
separately increase and decrease as the bandwidth becomes larger. It
is also detected from these tables that the poor estimates of
$\Sigma_{\alpha}(t,t)$'s appear at extremely small or large
bandwidths. In the numerical studies, the larger bias of
$\widehat{\theta}_{t}(\widehat{q}_{0.1 t})$ is found and the main
reason for this is because it is computed via comparing only (at
most) the top ten percent of subjects in the control group with
those in the case group. For $\alpha=0.3$, the availability of data
used for statistical analysis is expanded and, hence, the
performance becomes better. The biases of the proposed estimators is
indistinguishable in the presence of heavy censoring, whereas the
standard deviations and the standard errors will become larger. At
each simulated sample, we can see that the estimators using
$\lambda_{opt}$ provide generally satisfactory results.

The empirical coverage probabilities of pointwise confidence
intervals (tables 1-3) show the good performance of bandwidths
selected from the automatic selection procedure for interval
estimation, except for $(n,c.r.,\alpha)$ of $(500,50\%,0.1)$ and
$(500,50\%,0.2)$ at the time point $t_{0.6}$. However, most of the
coverage probabilities are lower than 0.95 for the bandwidths of
0.01 and 0.2. For samples without censoring, it is revealed from
tables 1-3 that the empirical coverage probabilities of the
pointwise confidence intervals computed based on
$\widehat{\theta}_{t}(\widehat{q}_{\alpha t})$ with $\lambda_{opt}$
are more close to the nominal level of 0.95 than those based on
$\widetilde{\theta}_{t}(\widetilde{q}_{\alpha t})$. Similar
conclusions can be also drawn from table 4 for the simultaneous
coverage probabilities. The empirical coverage probabilities of the
simultaneous confidence band (\ref{2.6}) with $\lambda_{opt}$
roughly stay around 0.95 except for the wider interval
$[t_{0.4},t_{0.6}]$.

\section{A Data Example - ACTG 175 Study}

\renewcommand{\theequation}{4.\arabic{equation}}
\setcounter{equation}{0}

In the ACTG 175 study (Hammer et al. (1996)), the classification
accuracy of CD4 cell counts on the time in weeks from entry to AIDS
diagnosis or death might depend on whether they received the prior
antiretroviral therapy. A total of 2467 HIV-1-infected patients,
which were recruited between December 1991 and October 1992, are
considered. Of these patients, 1395 received the prior therapy while
the rest 1072 did not receive the therapy. During the study period,
308 patients died of all causes or were diagnosed with AIDS. For a
negative association between CD4 counts and the time to AIDS and
death, we let $Y$ be a strictly decreasing function of the CD4
marker and $T$ be the minimum of time-to-AIDS and time-to-death.
Currently, there is still no standard of clinically meaningful
values of FPR for the pAUC in AIDS research. In this data analysis,
we restrict our attention to the pAUC of $Y$ with the FPR less than
0.1 or 0.2 or 0.3.

To simplify the presentation, the marker and the time-dependent pAUC
for non-therapy and therapy patients are denoted separately by
$(Y^{(1)},\theta_{t}^{(1)}(q^{(1)}_{\alpha t}))$ and
$(Y^{(2)},\theta_{t}^{(2)}(q^{(2)}_{\alpha t}))$. Based on two
independent data sets $\{X^{(1)}_i,\delta^{(1)}_i
,Y^{(1)}_i\}_{i=1}^{n_1}$ and $\{X^{(2)}_i,\delta^{(2)}_i,
Y^{(2)}_i\}_{i=1}^{n_2}$,
$\widehat{\theta}_{t}^{(1)}(\widehat{q}^{(1)}_{\alpha t})$ and
$\widehat{\theta}_{t}^{(2)}(\widehat{q}^{(2)}_{\alpha t})$ are
computed as $\widehat{\theta}_{t}(\widehat{q}_{\alpha t})$ in
(\ref{2.1}) using the bandwidths of 0.042 and 0.069, which are the
minimizers of $ISE(\lambda)$ in (\ref{3.1}). The confidence
intervals for $\theta_{t}^{(k)}(q^{(k)}_{\alpha t})$'s are further
constructed from (\ref{2.4}) and (\ref{2.6}). Due to the large
variation of estimators before week 98, we only provide the
estimated time-dependent pAUCs and 0.95 pointwise and simultaneous
confidence bands from week 98 to the end of study in figures 1
(a)-(f). Based on the summary measures
$\theta_{t}^{(k)}(q^{(k)}_{0.1 t})$'s and compared with 0.005, one
can conclude from the simultaneous confidence bands that the CD4
count is a useless biomarker in classifying patient's survival time
within the considered time period for both therapy and non-therapy
patients. However, a different conclusion will be drawn at each time
point based on the pointwise confidence intervals. The
time-dependent pAUCs $\theta_{t}^{(1)}(q^{(1)}_{\alpha t})$ and
$\theta_{t}^{(2)}(q^{(2)}_{\alpha t})$ are detected to be
significantly higher than $0.5\alpha^{2}$ for $\alpha$=0.2 and 0.3
after week 110 and week 98, respectively. Figures 1 (a), (c), and
(e) give a clear indication that the pAUCs decrease slightly over
time for patients without prior therapy. However, the pAUCs stay
very close to a constant throughout the study period for those with
prior therapy (figures 1 (b), (d), and (f)).

The difference in the classification accuracies of $Y^{(1)}$ and
$Y^{(2)}$ can be measured by the summary index
$\gamma_{\alpha}(t)=\theta_{t}^{(1)}(q^{(1)}_{\alpha
t})-\theta_{t}^{(2)}(q^{(2)}_{\alpha t})$. When $\alpha=1$,
$\gamma_{\alpha}(t)$ is the usual comparison of AUCs in the
time-dependent setting. It is natural to estimate
$\gamma_{\alpha}(t)$ by $
\widehat{\gamma}_{\alpha}(t)=\widehat{\theta}_{t}^{(1)}(\widehat{q}^{(1)}_{\alpha
t})-\widehat{\theta}_{t}^{(2)}(\widehat{q}^{(2)}_{\alpha t})$. Along
the same lines as the proof in the Appendix, we can derive that
$\sqrt{n}(\widehat{\gamma}_{\alpha}(t)-\gamma_{\alpha}(t))$
converges weakly to a mean zero Gaussian process with
variance-covariance function
$\Gamma_{\alpha}(s,t)=\kappa^{-1}E[\Psi_{\alpha
i}^{(1)}(s)\Psi_{\alpha i}^{(1)}(t)]+(1-\kappa)^{-1}E[\Psi_{\alpha
i}^{(2)}(s)\Psi_{\alpha i}^{(2)}(t)]$ provided that
$n_1/n\rightarrow\kappa$ ($0<\kappa<1$) as
$n=(n_{1}+n_{2})\rightarrow\infty$, where $\Psi_{\alpha i}^{(k)}(t)$
is a counterpart of $\Psi_{\alpha i}(t)$, $k=1,2$. To make inference
on $\gamma_{\alpha}(t)$, $\Gamma_{\alpha}(s,t)$ is first estimated
by
\begin{eqnarray}
\widehat{\Gamma}_{\alpha}(s,t)=\frac{n}{n_1^2}\sum_{i=1}^{n_1}\widehat{\Psi}_{\alpha
i}^{(1)}(s)\widehat{\Psi}_{\alpha
i}^{(1)}(t)+\frac{n}{n_2^2}\sum_{i=1}^{n_2}\widehat{\Psi}_{\alpha
i}^{(2)}(s)\widehat{\Psi}_{\alpha i}^{(2)}(t).\label{4.5}
\end{eqnarray}
A $(1-\varsigma)$ pointwise confidence interval for
$\gamma_{\alpha}(t)$ and a $(1-\varsigma)$ simultaneous confidence
band for $\{\gamma_{\alpha}(t): t\in[\tau_1,\tau_2]\}$ are
separately given via
\begin{eqnarray}
\widehat{\gamma}_{\alpha}(t)\pm\frac{Z_{\varsigma/2}}{\sqrt{n}}\widehat{\Gamma}_{\alpha}^{\frac{1}{2}}(t,t)
\text{ and }\{\widehat{\gamma}_{\alpha}(t)\pm\frac{L^{(\gamma)}
_{\varsigma}}{\sqrt{n}}\widehat{\Gamma}_{\alpha}^{\frac{1}{2}}(t,t)
:t\in[\tau_1,\tau_2]\}\label{4.6}
\end{eqnarray}
with $L^{(\gamma)}_{\varsigma}$ being obtained as (\ref{2.5}). It is
revealed in figures 2 (a)-(c) that $\gamma_{\alpha}(t)$,
$\alpha$=0.1, 0.2, and 0.3, tend to be positive within the study
period and the difference becomes negligible as $\alpha$ increases.
In other words, with small values of $FPR_t(y)$, a prior
antiretroviral therapy might lower the discrimination ability of CD4
counts in classifying subject's t-week survival. One possible
explanation for this conclusion is that the prior therapy makes
patients more homogeneous in survival time and CD4 counts. The
estimates are further found to be around zero after about week 160.
It means that for long term survival classification the performance
of CD4 counts is irrelevant to whether patients receive prior
therapy or not. Due to the large variability in the data, we could
not detect any significant difference between
$\theta_{t}^{(1)}(q^{(1)}_{\alpha t})$ and
$\theta_{t}^{(2)}(q^{(2)}_{\alpha t})$. It would necessitate
extremely large sample sizes to enable demonstration of significant
differences between the pAUCs.

\section{Discussion}

For the time-dependent pAUC, it was traditionally estimated by the
trapezoidal numerical integration method. The derivation for its
sampling distribution becomes complicated and the computation load
is prohibitively expensive. Although the inferences can be developed
through a bootstrap technique, there is still no rigorous
theoretical justification for this procedure. We can see in this
article that the proposed estimators are simple and have explicit
mathematical expressions. The confidence bands are built based on
the asymptotic Gaussian process of the estimators as well as the
corresponding estimates of the asymptotic variances. The estimation
and inference procedures are further shown to be useful through
simulation studies and an application to the ACTG 175 data.

It is detected from our numerical studies that the performance of
the proposed estimator $\widehat{\theta}_t(\widehat{q}_{\alpha t})$
is very sensitive to small value of $\alpha$. To obtain a more
stable estimate of $\theta_t(q_{\alpha t})$, a large sample size
relative to $\alpha$ is usually suggested especially in the presence
of censoring. Moreover, the price for the assumption of marker
dependent censoring is to find an appropriate bandwidth in
estimation. To this problem, we propose a simple and easily
implemented selection procedure and show its good performance
through simulations. As for the estimation of $\theta_t(q_{\alpha
t})$, it can be also derived via using the bivariate estimation
methods of Campbell (1981) or Burke (1988) for $S(t,y)$. However,
these estimators are only valid under independent censorship which
appears to be very limited and may not always be met in
applications. One advantage of totally independent censoring
assumption is that no smoothing technique is required.

In some empirical examples, censored survival data of the form
$\{X_i,\delta_i, Y^{(1)}_i,Y^{(2)}_i\}_{i=1}^n$ are often occurred
in a paired design with $(Y^{(1)}_{i},Y^{(2)}_{i})$ being the
different biomarkers of the $i$th subject. The scientific interest
usually focuses on comparing the discrimination abilities of
$Y^{(1)}$ and $Y^{(2)}$ on subject's survival status at each time
point within the study period. Obviously, the assumptions of marker
dependent censoring made separately on $(T,C,Y^{(1)})$ and
$(T,C,Y^{(2)})$ are often unreasonable in practice. Under a more
flexible assumption of conditionally independent censorship ($T$ and
$C$ are independent conditioning on $(Y^{(1)},Y^{(2)})$), the
estimated joint survivor function of $T$ and $Y^{(1)}$ and that of
$T$ and $Y^{(2)}$ in this article are quite inadequate in the
estimation of $\gamma_{\alpha}(t)$ without modification. It is
worthwhile to investigate the associated comparison procedure in our
future study.

\medskip

\begin{center}
{\large APPENDIX}
\end{center}
\renewcommand{\theequation}{A.\arabic{equation}} \setcounter{equation}{0}

\setcounter{section}{0}

For the proof of main results, the assumptions in Akritas (1994) and
the conditions (\textbf{A1}: $f_t(y)=-\partial FPR_t(y)/\partial y$
exists with $\inf_tf_t(q_{\alpha t})>0$) and (\textbf{A2}:
$\sup_t|\varsigma^{-1}\{FPR_t(q_{\alpha
t}+\varsigma)-FPR_t(q_{\alpha t})\}+f_t(q_{\alpha t})|\rightarrow 0$
as $\varsigma \rightarrow 0$) are made throughout the rest of this
paper.

\medskip

\noindent \textbf{Asymptotic Gaussian Process of
$\widehat{\theta}_t(\widehat{q}_{\alpha t})$:}

\medskip

From Theorem 3.1 of Akritas (1994), one has
\begin{eqnarray}
\sup_{t,y}|\sqrt{n}(\widehat{S}(t,y)-S(t,y))-\frac{\sqrt{n}}{n}\sum_{i=1}^nV_{i}(t,y)|=o_p(1),
\label{a.5}
\end{eqnarray}
where $V_{i}(t,y)=(S_T(t|Y_i)+\xi_i(t))I(Y_i>y)-S(t,y)$ with
$\xi_i(t)=-S_T(t|Y_i)\int_0^tS_X^{-1}(u|Y_i)$ $d_uM_i(u|Y_i)$,
$M_i(t|Y_i)=I(X_i \leq t)\delta_i+\ln S_{T}(t\wedge X_i|Y_i)$, and
$t\wedge X_i=\min\{t,X_{i}\}$. Let
$h_{ij}(t,y)=(1-S_T(t|Y_i))S_T(t|Y_j)\phi_{ij}(y)$ and
$H(t,y)=E[h_{ij}(t,y)]$. The uniform consistency of
$\widehat{S}_T(t|y)$ (cf. Dabrowska (1987)) ensures that
\begin{eqnarray}
\widehat{H}(t,y)&=&\frac{1}{n^2}\sum_{i\neq
j}h_{ij}(t,y)+\frac{1}{n^2}\sum_{i\neq
j}(S_T(t|Y_i)-\widehat{S}_T(t|Y_i))S_T(t|Y_j)\phi_{ij}(y)\nonumber\\&&+\frac{1}{n^2}\sum_{i\neq
j}(1-S_T(t|Y_i))(\widehat{S}_T(t|Y_j)-S_T(t|Y_j))\phi_{ij}(y)+r_{1n}(t,y)
\label{a.1}
\end{eqnarray}
with $\sup_{t,y}|r_{1n}(t,y)|=o_p(n^{-1/2})$. By a direct
calculation and (\ref{a.5}), a simplified form of the second term in
the righthand side of (\ref{a.1}) is obtained as follows:
\begin{eqnarray}
\lefteqn{\frac{1}{n}\sum_{j=1}^nS_T(t|Y_j)I(Y_j>y)\{\frac{1}{n}\sum_{i=1}^nS_T(t|Y_i)I(Y_i>Y_j)-\widehat{S}(t,Y_j)\}}\nonumber\\
&=&\frac{-1}{n^2}\sum_{i,j}S_T(t|Y_j)\xi_i(t)\phi_{ij}(y)+r_{2n}(t,y),
\label{a.2}
\end{eqnarray}
where $\sup_{t,y}|r_{2n}(t,y)|=o_p(n^{-1/2})$. Similarly, the third
term can be expressed as
\begin{eqnarray}
\frac{1}{n^2}\sum_{i,j}(1-S_T(t|Y_i))\xi_j(t)\phi_{ij}(y)+r_{3n}(t,y)
\label{a.3}
\end{eqnarray}
with $\sup_{t,y}|r_{3n}(t,y)|=o_p(n^{-1/2})$. It follows from
(\ref{a.1})-(\ref{a.3}), the decomposition of a U-statistic into a
sum of degenerate U-statistics (Serfling (1980)), and Corollary 4 of
Sherman (1994) that
\begin{eqnarray}
\sup_{t,y}|\sqrt{n}(\widehat{H}(t,y)-H(t,y))-\frac{\sqrt{n}}{n}\sum_{i=1}^nU_{i}(t,y)|=o_p(1),
\label{a.4}
\end{eqnarray}
where
$U_{i}(t,y)=E[h_{ij}(t,y)+h_{ji}(t,y)|X_i,Y_i,\delta_i]-2H(t,y)
+(S_Y(Y_i)-S(t,y))\xi_i(t)I(Y_i>y)$. By the Taylor expansion of
$\widehat{\theta}_t(y)=\widehat{H}(t,y)\{\widehat{S}_T(t)(1-\widehat{S}_T(t))\}^{-1}$
at $(\widehat{H}(t,y),\widehat{S}_{T}(t))=(H(t,y),S_{T}(t))$,
(\ref{a.5}), and (\ref{a.4}), one has
\begin{equation}
\sup_{t,y}|\sqrt{n}(\widehat{\theta}_{t}(y)-\theta_{t}(y))-\frac{\sqrt{n}}{n}\sum_{i=1}^n\frac{U_{i}(t,y)+
\eta(t,y)V_{i}(t,-\infty)}{S_T(t)(1-S_T(t))}|=o_p(1) \label{a.6},
\end{equation}
where $\eta(t,y)=H(t,y)(2S_T(t)-1)(S_T(t)-S_T^2(t))^{-1}$. Thus,
$\sqrt{n}(\widehat{\theta}_{t}(q_{\alpha t})-\theta_{t}(q_{\alpha
t}))$ can be shown to converge to a mean zero Gaussian process by an
application of the functional cental limit theorem.

For the asymptotic Gaussian process of
$\widehat{\theta}_t(\widehat{q}_{\alpha t})$, it is established
through the equality
$\sqrt{n}(\widehat{\theta}_t(\widehat{q}_{\alpha
t})-\theta_t(q_{\alpha
t}))=\sqrt{n}(\widehat{\theta}_t(\widehat{q}_{\alpha
t})-\theta_t(\widehat{q}_{\alpha t}))
+\sqrt{n}(\theta_t(\widehat{q}_{\alpha t})-\theta_t(q_{\alpha t}))$.
Let $\sqrt{n}(\widehat{q}_{\alpha t}-q_{\alpha
t})=\sqrt{n}(Q(\widehat{S})-Q(S))$ with $Q:S\rightarrow q_{\alpha
t}$. By assumptions (A1)-(A2), the Hadamard differentiability of $Q$
is a direct result of Lemma A.1 in Daouia, Florens, and Simar
(2008). Together with the functional delta method (cf. Van der Vaart
(2000)), we have
\begin{equation}
\sup_{t}|\sqrt{n}(\widehat{q}_{\alpha t}-q_{\alpha
t})-\frac{\sqrt{n}}{n}\sum_{i=1}^n\frac{V_{i}(t,q_{\alpha t})-\alpha
V_{i}(t,-\infty)} {f_t(q_{\alpha t})S_T(t)}|=o_p(1) \label{a.8}.
\end{equation}
and the weak convergence of $\sqrt{n}(\widehat{q}_{\alpha
t}-q_{\alpha t})$. It is further ensured by (a version of) Lemma
19.24 of van der Vaart (2000) and (\ref{a.6}) that
\begin{equation}
\sup_t|\sqrt{n}(\widehat{\theta}_{t}(\widehat{q}_{\alpha
t})-\theta_{t}(\widehat{q}_{\alpha
t})-\sqrt{n}(\widehat{\theta}_{t}(q_{\alpha t})-\theta_{t}(q_{\alpha
t}))|=o_p(1). \label{a.12}
\end{equation}
Moreover, the first order Taylor expansion of
$\theta_t(\widehat{q}_{\alpha t})$ at $\widehat{q}_{\alpha
t}=q_{\alpha t}$, the continuity of $\partial \theta_t(y)/\partial
y$, $\sup_t|\widehat{q}_{\alpha t}-q_{\alpha t}|=o_p(1)$, and the
continuous mapping theorem imply that
\begin{equation}
\sup_t|\sqrt{n}(\theta_t(\widehat{q}_{\alpha t})-\theta_t(q_{\alpha
t}))-\frac{\alpha S_T(t)-S_Y(q_{\alpha t})}{1-S_T(t)}f_t(q_{\alpha
t})\sqrt{n}(\widehat{q}_{\alpha t}-q_{\alpha t})|=o_p(1).
\label{a.10}
\end{equation}
It follows from (\ref{a.6})-(\ref{a.10}) that
\begin{equation}
\sup_t|\sqrt{n}(\widehat{\theta}_t(\widehat{q}_{\alpha
t})-\theta_t(q_{\alpha t}))-
\frac{\sqrt{n}}{n}\sum_{i=1}^n\Psi_{\alpha i}(t)|=o_p(1),
\label{a.11}
\end{equation}
\begin{equation}
\text{where } \Psi_{\alpha i}(t)=\frac{U_{i}(t,q_{\alpha
t})+\eta(t,q_{\alpha t})V_{i}(t,-\infty)+ (\alpha
S_T(t)-S_Y(q_{\alpha t}))(V_{i}(t,q_{\alpha t})-\alpha
V_{i}(t,-\infty))}{S_T(t)(1-S_T(t))}\nonumber.
\end{equation}
Finally, the proof is completed by applying the functional central
limit theorem to the approximated term
$n^{-1/2}\sum_{i=1}^n\Psi_{\alpha i}(t)$ in (\ref{a.11}).

\medskip

\begin{center}
{\large REFERENCES}
\end{center}

\begin{description}
\item[] Akritas, M. G. (1994). Nearest neighbor estimation of a
bivariate distribution under random censoring. {\em Annals of
Statistics} {\bf {22}}, 1299-1327.
\item[] Burke, M. D. (1988). Estimation of a bivariate distribution
function under random censorship. {\em Biometrika} {\bf {75}},
379-382.
\item[] Cai, T. and Dodd, L. E. (2008). Regression analysis for the
partial area under the ROC curve. {\em Statistica Sinica} {\bf
{18}}, 817-836.
\item[] Campbell, G. (1981). Nonparametric bivariate estimation with
randomly censored data. {\em Biometrika} {\bf {68}}, 417-423.
\item[] Chambless, L. E. and Diao, G. (2006). Estimation of time-dependent area under the ROC curve for long-term risk prediction.
{\em Statistics in Medicine} {\bf{25}}, 3474-3486.
\item[] Chiang, C. T., Wang, S. H., and Hung, H. (2008). Random
weighting and Edgeworth expansion for the nonparametric
time-dependent AUC estimator. {\em Statistic Sinica}. Accepted.
\item[] Chiang, C. T. and Hung, H. (2008). Nonparametric estimation
for time-dependent AUC. Technical Report, National Taiwan
University.
\item[] Dabrowska, D. M. (1987). Uniform consistency of nearest
neighbor and kernel conditional Kaplan-Meier estimates. Technical
Report No. 86, Univ. California, Berkeley.
\item[] Daouia, A., Florens, J. P., and Simar, L. (2008). Functional convergence of
quantile-type frontiers with application to parametric
approximations. {\em Journal of Statistical Planning and Inference}
{\bf {138}}, 708-725.
\item[] Dodd, L. and Pepe, M. S. (2003). Partial AUC estimation and
regression. {\em Biometrics} {\bf{59}}, 614-623.
\item[] Dwyer, A. J. (1996). In pursuit of a piece of the ROC. {\em Radiology}
{\bf {201}}, 621-625.
\item[] Emir, B., Wieand, S., Jung, S. H., and Ying, Z. (2000).
Comparison of diagnostic markers with repeated measurements: a
non-parametric ROC curve approach. {\em Statistics in Medicine} {\bf
{19}}, 511-523.
\item[] Hammer, S. M., Katzenstein, D. A., Hughes, M. D., Gundacker, H., Schooley, R. T., Haubrich, R. H.,
Henry, W. K., Lederman, M. M., Phair, J. P., Niu, M., Hirsch, M. S.,
and Merigan, T. C. (1996). A trial comparing nucleoside monotherapy
with combination therapy in HIV-infected adults with CD4 cell counts
from 200 to 500 per cubic millimeter. {\em The New England Journal
of Medicine} {\bf {335}}, 1081-1090.
\item[] Heagerty, P. J., Lumley, T., and Pepe, M. S. (2000).
Time-dependent ROC curves for censored survival data and a
diagnostic marker. {\em Biometrics} {\bf {54}}, 124-135.
\item[] Jiang, Y., Metz, C. E., and Nishikawa, R. M. (1996).
A receiver operating characteristic partial area index for highly
sensitive diagnostic tests. {\em Radiology} {\bf {201}}, 745-750.
\item[] Lin, D. Y., Wei, L. J., Yang, I., and Ying, Z. (2000).
Semiparametric regression for the mean and rate functions of
recurrent events. {\em Journal of the Royal Statistical Society}
{\bf{B62}}, 711-730.
\item[] McClish, D. K. (1989). Analyzing a portion of the ROC curve.
{\em Medical Decision Making} {\bf {9}}, 190-195.
\item[] Serfling, R. J. (1980). Approximation theorems of
mathematical statistics. New York: Wiley.
\item[] Sherman, R. P. (1994). Maximal inequalities for degenerate U-processes
with applications to optimization estimators. {\em Annals of
Statistics} {\bf {22}}, 439-459.
\item[] van der Vaart, A. W. (2000). Asymptotic Statistics. Cambridge: Cambridge University Press.
\item[] Zhang, D. D., Zhou, X. H., Freeman, Daniel H. J., and
Freeman, J. L. (2002). A non-parametric method for the comparison of
partial areas under ROC curves and its application to large health
care data set. {\em Statistics in Medicine} {\bf {21}}, 701-715.
\item[] Zhou, X. H., McClish, D. K., and
Obuchowski, N. A. (2002). Statistical methods in diagnostic
medicine. New York: Wiley.
\end{description}

\clearpage

\begin{center} \centerline{Table 1 } \small{ \emph{The averages $(Mean)$ and the standard deviations $(SD)$ of
500 estimates, the standard errors $(SE)$, and the empirical
coverage probabilities $(CP)$}} \vspace{0.1cm}

\tabcolsep=5pt
\begin{tabular}{ccccccccccc}
\hline \hline
$c.r.=0\%$  &       &       &   $n=500$ &       &       &       &   $n=1000$    &       &       &       \\
\hline
    &   Time    &   $\theta_t(q_{0.1 t})$   &   Mean    &   SD  &   SE  &   CP  &   Mean    &   SD  &   SE  &   CP  \\
\hline
    &   $t_{0.4}$   &   0.0264  &   0.0273  &   0.0044  &   0.0041  &   0.938   &   0.0269  &   0.0031  &   0.0029  &   0.944   \\
$\widetilde{\theta}_t(\widetilde{q}_{0.1 t})$    &   $t_{0.5}$   &   0.0269  &   0.0284  &   0.0045  &   0.0042  &   0.918   &   0.0275  &   0.0031  &   0.0030  &   0.934   \\
    &   $t_{0.6}$   &   0.0279  &   0.0296  &   0.0046  &   0.0045  &   0.914   &   0.0288  &   0.0034  &   0.0031  &   0.926   \\
\hline
    &   $t_{0.4}$   &   0.0264  &   0.0240  &   0.0043  &   0.0049  &   0.954   &   0.0247  &   0.0030  &   0.0034  &   0.948   \\
$\widehat{\theta}_t(\widehat{q}_{0.1 t})$   & $t_{0.5}$   &   0.0269  &   0.0249  &   0.0042  &   0.0047  &   0.960   &   0.0256  &   0.0030  &   0.0032  &   0.938   \\
  $(\lambda_{opt})$  &   $t_{0.6}$   &   0.0279  &   0.0262  &   0.0046  &   0.0046  &   0.928   &   0.0269  &   0.0033  &   0.0033  &   0.942   \\
   &               &       &       &       &       &       &       &       &       \\
\hline \hline
$c.r.=30\%$ &       &       &   $n=500$ &       &       &       &   $n=1000$    &       &       &       \\
\hline
$\lambda$   &   Time    &   $\theta_t(q_{0.1 t})$   &   Mean    &   SD  &   SE  &   CP  &   Mean    &   SD  &   SE  &   CP  \\
\hline
    &   $t_{0.4}$   &   0.0264  &   0.0268  &   0.0052  &   0.0040  &   0.834   &   0.0264  &   0.0034  &   0.0030  &   0.908   \\
0.01    &   $t_{0.5}$   &   0.0269  &   0.0272  &   0.0055  &   0.0040  &   0.846   &   0.0272  &   0.0038  &   0.0030  &   0.874   \\
    &   $t_{0.6}$   &   0.0279  &   0.0279  &   0.0055  &   0.0041  &   0.848   &   0.0280  &   0.0040  &   0.0031  &   0.882   \\
\hline
    &   $t_{0.4}$   &   0.0264  &   0.0234  &   0.0050  &   0.0058  &   0.948   &   0.0243  &   0.0035  &   0.0041  &   0.950   \\
$\lambda_{opt}$ &   $t_{0.5}$   &   0.0269  &   0.0248  &   0.0054  &   0.0056  &   0.932   &   0.0252  &   0.0037  &   0.0040  &   0.942   \\
    &   $t_{0.6}$   &   0.0279  &   0.0264  &   0.0055  &   0.0056  &   0.924   &   0.0264  &   0.0039  &   0.0040  &   0.910   \\
\hline
    &   $t_{0.4}$   &   0.0264  &   0.0190  &   0.0032  &   0.0062  &   0.894   &   0.0187  &   0.0021  &   0.0045  &   0.654   \\
0.20    &   $t_{0.5}$   &   0.0269  &   0.0203  &   0.0037  &   0.0061  &   0.868   &   0.0200  &   0.0026  &   0.0044  &   0.698   \\
    &   $t_{0.6}$   &   0.0279  &   0.0218  &   0.0045  &   0.0061  &   0.864   &   0.0218  &   0.0031  &   0.0045  &   0.762   \\
    &               &       &       &       &       &       &       &       &       \\
\hline \hline
$c.r.=50\%$ &       &       &   $n=500$ &       &       &       &   $n=1000$    &       &       &       \\
\hline
$\lambda$   &   Time    &   $\theta_t(q_{0.1 t})$   &   Mean    &   SD  &   SE  &   CP  &   Mean    &   SD  &   SE  &   CP  \\
\hline
    &   $t_{0.4}$   &   0.0264  &   0.0266  &   0.0062  &   0.0042  &   0.786   &   0.0266  &   0.0043  &   0.0033  &   0.864   \\
0.01    &   $t_{0.5}$   &   0.0269  &   0.0268  &   0.0064  &   0.0041  &   0.772   &   0.0270  &   0.0046  &   0.0033  &   0.828   \\
    &   $t_{0.6}$   &   0.0279  &   0.0265  &   0.0066  &   0.0043  &   0.754   &   0.0275  &   0.0051  &   0.0034  &   0.792   \\
\hline
    &   $t_{0.4}$   &   0.0264  &   0.0237  &   0.0057  &   0.0067  &   0.950   &   0.0243  &   0.0040  &   0.0047  &   0.960   \\
$\lambda_{opt}$ &   $t_{0.5}$   &   0.0269  &   0.0250  &   0.0061  &   0.0064  &   0.918   &   0.0253  &   0.0045  &   0.0046  &   0.924   \\
    &   $t_{0.6}$   &   0.0279  &   0.0266  &   0.0071  &   0.0064  &   0.892   &   0.0267  &   0.0049  &   0.0047  &   0.920   \\
\hline
    &   $t_{0.4}$   &   0.0264  &   0.0192  &   0.0037  &   0.0073  &   0.952   &   0.0190  &   0.0025  &   0.0053  &   0.828   \\
0.20    &   $t_{0.5}$   &   0.0269  &   0.0209  &   0.0047  &   0.0071  &   0.932   &   0.0203  &   0.0031  &   0.0052  &   0.846   \\
    &   $t_{0.6}$   &   0.0279  &   0.0225  &   0.0056  &   0.0072  &   0.908   &   0.0221  &   0.0037  &   0.0055  &   0.864   \\
    &               &       &       &       &       &       &       &       &       \\
\end{tabular}
\end{center}

\clearpage

\begin{center} \centerline{Table 2} \small{ \emph{The averages $(Mean)$ and the standard deviations $(SD)$ of
500 estimates, the standard errors $(SE)$, and the empirical
coverage probabilities $(CP)$}} \vspace{0.1cm}

\tabcolsep=5pt
\begin{tabular}{ccccccccccc}
\hline \hline
$c.r.=0\%$  &       &       &   $n=500$ &       &       &       &   $n=1000$    &       &       &       \\
\hline
    &   Time    &   $\theta_t(q_{0.2 t})$   &   Mean    &   SD  &   SE  &   CP  &   Mean    &   SD  &   SE  &   CP  \\
\hline
    &   $t_{0.4}$   &   0.0770  &   0.0786  &   0.0086  &   0.0083  &   0.930   &   0.0780  &   0.0062  &   0.0058  &   0.922   \\
$\widetilde{\theta}_t(\widetilde{q}_{0.2 t})$    &   $t_{0.5}$   &   0.0776  &   0.0801  &   0.0084  &   0.0082  &   0.924   &   0.0785  &   0.0059  &   0.0058  &   0.932   \\
    &   $t_{0.6}$   &   0.0794  &   0.0825  &   0.0086  &   0.0086  &   0.918   &   0.0808  &   0.0063  &   0.0060  &   0.920   \\
\hline
    &   $t_{0.4}$   &   0.0770  &   0.0732  &   0.0085  &   0.0093  &   0.956   &   0.0745  &   0.0060  &   0.0064  &   0.950   \\
$\widehat{\theta}_t(\widehat{q}_{0.2 t})$ &   $t_{0.5}$   &   0.0776  &   0.0744  &   0.0081  &   0.0089  &   0.966   &   0.0755  &   0.0058  &   0.0061  &   0.952   \\
 $(\lambda_{opt})$   &   $t_{0.6}$   &   0.0794  &   0.0765  &   0.0088  &   0.0088  &   0.936   &   0.0777  &   0.0061  &   0.0062  &   0.948   \\
   &               &       &       &       &       &       &       &       &       \\
\hline \hline
$c.r.=30\%$ &       &       &   $n=500$ &       &       &       &   $n=1000$    &       &       &       \\
\hline
$\lambda$   &   Time    &   $\theta_t(q_{0.2 t})$   &   Mean    &   SD  &   SE  &   CP  &   Mean    &   SD  &   SE  &   CP  \\
\hline
    &   $t_{0.4}$   &   0.0770  &   0.0774  &   0.0101  &   0.0080  &   0.850   &   0.0767  &   0.0067  &   0.0061  &   0.926   \\
0.01    &   $t_{0.5}$   &   0.0776  &   0.0779  &   0.0101  &   0.0078  &   0.864   &   0.0780  &   0.0072  &   0.0060  &   0.884   \\
    &   $t_{0.6}$   &   0.0794  &   0.0789  &   0.0100  &   0.0080  &   0.862   &   0.0797  &   0.0075  &   0.0061  &   0.874   \\
\hline
    &   $t_{0.4}$   &   0.0770  &   0.0724  &   0.0102  &   0.0111  &   0.944   &   0.0742  &   0.0070  &   0.0077  &   0.962   \\
$\lambda_{opt}$ &   $t_{0.5}$   &   0.0776  &   0.0744  &   0.0107  &   0.0106  &   0.926   &   0.0751  &   0.0071  &   0.0074  &   0.952   \\
    &   $t_{0.6}$   &   0.0794  &   0.0771  &   0.0104  &   0.0106  &   0.938   &   0.0770  &   0.0074  &   0.0075  &   0.946   \\
\hline
    &   $t_{0.4}$   &   0.0770  &   0.0648  &   0.0079  &   0.0125  &   0.932   &   0.0642  &   0.0053  &   0.0089  &   0.796   \\
0.20    &   $t_{0.5}$   &   0.0776  &   0.0667  &   0.0086  &   0.0119  &   0.914   &   0.0663  &   0.0061  &   0.0085  &   0.808   \\
    &   $t_{0.6}$   &   0.0794  &   0.0693  &   0.0095  &   0.0118  &   0.912   &   0.0694  &   0.0066  &   0.0086  &   0.854   \\
    &               &       &       &       &       &       &       &       &       \\
\hline \hline
$c.r.=50\%$ &       &       &   $n=500$ &       &       &       &   $n=1000$    &       &       &       \\
\hline
$\lambda$   &   Time    &   $\theta_t(q_{0.2 t})$   &   Mean    &   SD  &   SE  &   CP  &   Mean    &   SD  &   SE  &   CP  \\
\hline
    &   $t_{0.4}$   &   0.0770  &   0.0770  &   0.0120  &   0.0084  &   0.820   &   0.0771  &   0.0083  &   0.0066  &   0.880   \\
0.01    &   $t_{0.5}$   &   0.0776  &   0.0768  &   0.0122  &   0.0082  &   0.794   &   0.0775  &   0.0088  &   0.0065  &   0.846   \\
    &   $t_{0.6}$   &   0.0794  &   0.0759  &   0.0127  &   0.0084  &   0.760   &   0.0785  &   0.0097  &   0.0066  &   0.814   \\
\hline
    &   $t_{0.4}$   &   0.0770  &   0.0728  &   0.0118  &   0.0128  &   0.944   &   0.0740  &   0.0081  &   0.0089  &   0.966   \\
$\lambda_{opt}$ &   $t_{0.5}$   &   0.0776  &   0.0748  &   0.0123  &   0.0122  &   0.926   &   0.0753  &   0.0086  &   0.0086  &   0.940   \\
    &   $t_{0.6}$   &   0.0794  &   0.0774  &   0.0134  &   0.0122  &   0.902   &   0.0775  &   0.0090  &   0.0088  &   0.924   \\
\hline
    &   $t_{0.4}$   &   0.0770  &   0.0654  &   0.0090  &   0.0146  &   0.966   &   0.0650  &   0.0064  &   0.0105  &   0.900   \\
0.20    &   $t_{0.5}$   &   0.0776  &   0.0679  &   0.0105  &   0.0138  &   0.950   &   0.0670  &   0.0072  &   0.0101  &   0.906   \\
    &   $t_{0.6}$   &   0.0794  &   0.0704  &   0.0115  &   0.0138  &   0.930   &   0.0700  &   0.0078  &   0.0102  &   0.908   \\
    &               &       &       &       &       &       &       &       &       \\

\end{tabular}
\end{center}

\clearpage

\begin{center} \centerline{Table 3 } \small{ \emph{The averages $(Mean)$ and the standard deviations $(SD)$ of
500 estimates, the standard errors $(SE)$, and the empirical
coverage probabilities $(CP)$}} \vspace{0.1cm}

\tabcolsep=5pt
\begin{tabular}{ccccccccccc}
\hline \hline
$c.r.=0\%$  &       &       &   $n=500$ &       &       &       &   $n=1000$    &       &       &       \\
\hline
    &   Time    &   $\theta_t(q_{0.3 t})$   &   Mean    &   SD  &   SE  &   CP  &   Mean    &   SD  &   SE  &   CP  \\
\hline
    &   $t_{0.4}$   &   0.1420  &   0.1442  &   0.0120  &   0.0118  &   0.944   &   0.1433  &   0.0088  &   0.0083  &   0.940   \\
$\widetilde{\theta}_t(\widetilde{q}_{0.3 t})$    &   $t_{0.5}$   &   0.1423  &   0.1457  &   0.0115  &   0.0116  &   0.938   &   0.1434  &   0.0083  &   0.0082  &   0.936   \\
    &   $t_{0.6}$   &   0.1446  &   0.1483  &   0.0120  &   0.0119  &   0.930   &   0.1461  &   0.0085  &   0.0084  &   0.938   \\
\hline
    &   $t_{0.4}$   &   0.1420  &   0.1372  &   0.0121  &   0.0131  &   0.962   &   0.1388  &   0.0084  &   0.0090  &   0.952   \\
$\widehat{\theta}_t(\widehat{q}_{0.3 t})$ &   $t_{0.5}$   &   0.1423  &   0.1383  &   0.0117  &   0.0124  &   0.954   &   0.1396  &   0.0080  &   0.0086  &   0.958   \\
  $(\lambda_{opt})$  &   $t_{0.6}$   &   0.1446  &   0.1408  &   0.0123  &   0.0124  &   0.944   &   0.1424  &   0.0083  &   0.0086  &   0.962   \\
   &               &       &       &       &       &       &       &       &       \\
\hline \hline
$c.r.=30\%$ &       &       &   $n=500$ &       &       &       &   $n=1000$    &       &       &       \\
\hline
$\lambda$   &   Time    &   $\theta_t(q_{0.3 t})$   &   Mean    &   SD  &   SE  &   CP  &   Mean    &   SD  &   SE  &   CP  \\
\hline
    &   $t_{0.4}$   &   0.1420  &   0.1424  &   0.0140  &   0.0116  &   0.880   &   0.1417  &   0.0095  &   0.0088  &   0.940   \\
0.01    &   $t_{0.5}$   &   0.1423  &   0.1427  &   0.0142  &   0.0113  &   0.874   &   0.1430  &   0.0101  &   0.0085  &   0.882   \\
    &   $t_{0.6}$   &   0.1446  &   0.1438  &   0.0137  &   0.0115  &   0.892   &   0.1451  &   0.0105  &   0.0086  &   0.874   \\
\hline
    &   $t_{0.4}$   &   0.1420  &   0.1362  &   0.0144  &   0.0154  &   0.948   &   0.1387  &   0.0098  &   0.0106  &   0.962   \\
$\lambda_{opt}$ &   $t_{0.5}$   &   0.1423  &   0.1383  &   0.0150  &   0.0147  &   0.932   &   0.1392  &   0.0099  &   0.0103  &   0.956   \\
    &   $t_{0.6}$   &   0.1446  &   0.1414  &   0.0143  &   0.0146  &   0.940   &   0.1415  &   0.0102  &   0.0104  &   0.948   \\
\hline
    &   $t_{0.4}$   &   0.1420  &   0.1265  &   0.0122  &   0.0174  &   0.938   &   0.1257  &   0.0082  &   0.0124  &   0.844   \\
0.20    &   $t_{0.5}$   &   0.1423  &   0.1281  &   0.0129  &   0.0164  &   0.914   &   0.1276  &   0.0090  &   0.0118  &   0.836   \\
    &   $t_{0.6}$   &   0.1446  &   0.1310  &   0.0137  &   0.0164  &   0.916   &   0.1313  &   0.0093  &   0.0118  &   0.874   \\
    &               &       &       &       &       &       &       &       &       \\
\hline \hline
$c.r.=50\%$ &       &       &   $n=500$ &       &       &       &   $n=1000$    &       &       &       \\
\hline
$\lambda$   &   Time    &   $\theta_t(q_{0.3 t})$   &   Mean    &   SD  &   SE  &   CP  &   Mean    &   SD  &   SE  &   CP  \\
\hline
    &   $t_{0.4}$   &   0.1420  &   0.1417  &   0.0167  &   0.0122  &   0.826   &   0.1419  &   0.0116  &   0.0095  &   0.892   \\
0.01    &   $t_{0.5}$   &   0.1423  &   0.1407  &   0.0168  &   0.0118  &   0.824   &   0.1420  &   0.0120  &   0.0093  &   0.866   \\
    &   $t_{0.6}$   &   0.1446  &   0.1389  &   0.0176  &   0.0120  &   0.796   &   0.1432  &   0.0133  &   0.0095  &   0.826   \\
\hline
    &   $t_{0.4}$   &   0.1420  &   0.1369  &   0.0168  &   0.0176  &   0.948   &   0.1383  &   0.0115  &   0.0122  &   0.964   \\
$\lambda_{opt}$ &   $t_{0.5}$   &   0.1423  &   0.1390  &   0.0173  &   0.0168  &   0.934   &   0.1394  &   0.0119  &   0.0119  &   0.948   \\
    &   $t_{0.6}$   &   0.1446  &   0.1418  &   0.0184  &   0.0169  &   0.928   &   0.1421  &   0.0122  &   0.0121  &   0.948   \\
\hline
    &   $t_{0.4}$   &   0.1420  &   0.1273  &   0.0137  &   0.0201  &   0.968   &   0.1267  &   0.0098  &   0.0144  &   0.910   \\
0.20    &   $t_{0.5}$   &   0.1423  &   0.1295  &   0.0154  &   0.0190  &   0.954   &   0.1285  &   0.0105  &   0.0138  &   0.910   \\
    &   $t_{0.6}$   &   0.1446  &   0.1321  &   0.0163  &   0.0190  &   0.938   &   0.1318  &   0.0113  &   0.0139  &   0.904   \\
    &               &       &       &       &       &       &       &       &       \\

\end{tabular}
\end{center}

\clearpage

\begin{center} \centerline{Table 4 } \small{ \emph{The empirical
coverage probabilities of $0.95$ simultaneous confidence
bands}}\vspace{0.1cm}

\tabcolsep=5pt
\begin{tabular}{cccccc}
\hline \hline
$c.r.=0\%$  &       &   $n=500$ &       &   $n=1000$    &       \\
\hline
    &   $\alpha$    &   $[t_{0.4},t_{0.5}]$ &   $[t_{0.4},t_{0.6}]$ &   $[t_{0.4},t_{0.5}]$ &   $[t_{0.4},t_{0.6}]$ \\
\hline
    &   0.1 &   0.904   &   0.892   &   0.932   &   0.928   \\
$\widetilde{\theta}_t(\widetilde{q}_{\alpha t})$    &   0.2 &   0.922   &   0.908   &   0.932   &   0.934   \\
    &   0.3 &   0.930   &   0.908   &   0.940   &   0.942   \\
\hline
    &   0.1 &   0.962   &   0.942   &   0.936   &   0.932   \\
$\widehat{\theta}_t(\widehat{q}_{\alpha t})$ &   0.2 &   0.968   &   0.952   &   0.948   &   0.944   \\
 $\lambda_{opt}$   &   0.3 &   0.958   &   0.950   &   0.946   &   0.952   \\
    &       &       &       &       &       \\
     \hline \hline
$c.r.=30\%$ &       &   $n=500$ &       &   $n=1000$    &       \\
\hline
$\lambda$   &   $\alpha$    &   $[t_{0.4},t_{0.5}]$ &   $[t_{0.4},t_{0.6}]$ &   $[t_{0.4},t_{0.5}]$ &   $[t_{0.4},t_{0.6}]$ \\
\hline
    &   0.1 &   0.800   &   0.754   &   0.854   &   0.824   \\
0.01    &   0.2 &   0.838   &   0.802   &   0.890   &   0.862   \\
    &   0.3 &   0.860   &   0.838   &   0.892   &   0.856   \\
\hline
    &   0.1 &   0.936   &   0.920   &   0.948   &   0.924   \\
$\lambda_{opt}$ &   0.2 &   0.942   &   0.928   &   0.952   &   0.948   \\
    &   0.3 &   0.938   &   0.938   &   0.952   &   0.952   \\
\hline
    &   0.1 &   0.888   &   0.878   &   0.708   &   0.732   \\
0.20    &   0.2 &   0.920   &   0.924   &   0.830   &   0.856   \\
    &   0.3 &   0.934   &   0.936   &   0.846   &   0.876   \\
    &       &       &       &       &       \\
\hline \hline
$c.r.=50\%$ &       &   $n=500$ &       &   $n=1000$    &       \\
\hline
$\lambda$   &   $\alpha$    &   $[t_{0.4},t_{0.5}]$ &   $[t_{0.4},t_{0.6}]$ &   $[t_{0.4},t_{0.5}]$ &   $[t_{0.4},t_{0.6}]$ \\
\hline
    &   0.1 &   0.702   &   0.614   &   0.778   &   0.726   \\
0.01    &   0.2 &   0.750   &   0.688   &   0.808   &   0.756   \\
    &   0.3 &   0.774   &   0.720   &   0.830   &   0.798   \\
\hline
    &   0.1 &   0.924   &   0.886   &   0.954   &   0.926   \\
$\lambda_{opt}$ &   0.2 &   0.936   &   0.898   &   0.954   &   0.936   \\
    &   0.3 &   0.948   &   0.914   &   0.946   &   0.942   \\
\hline
    &   0.1 &   0.940   &   0.896   &   0.868   &   0.860   \\
0.20    &   0.2 &   0.966   &   0.948   &   0.918   &   0.908   \\
    &   0.3 &   0.966   &   0.952   &   0.918   &   0.918   \\
    &       &       &       &       &       \\
\end{tabular}
\end{center}

\clearpage

\begin{figure}
\centering
    \includegraphics[width=6in,height=5.5in]{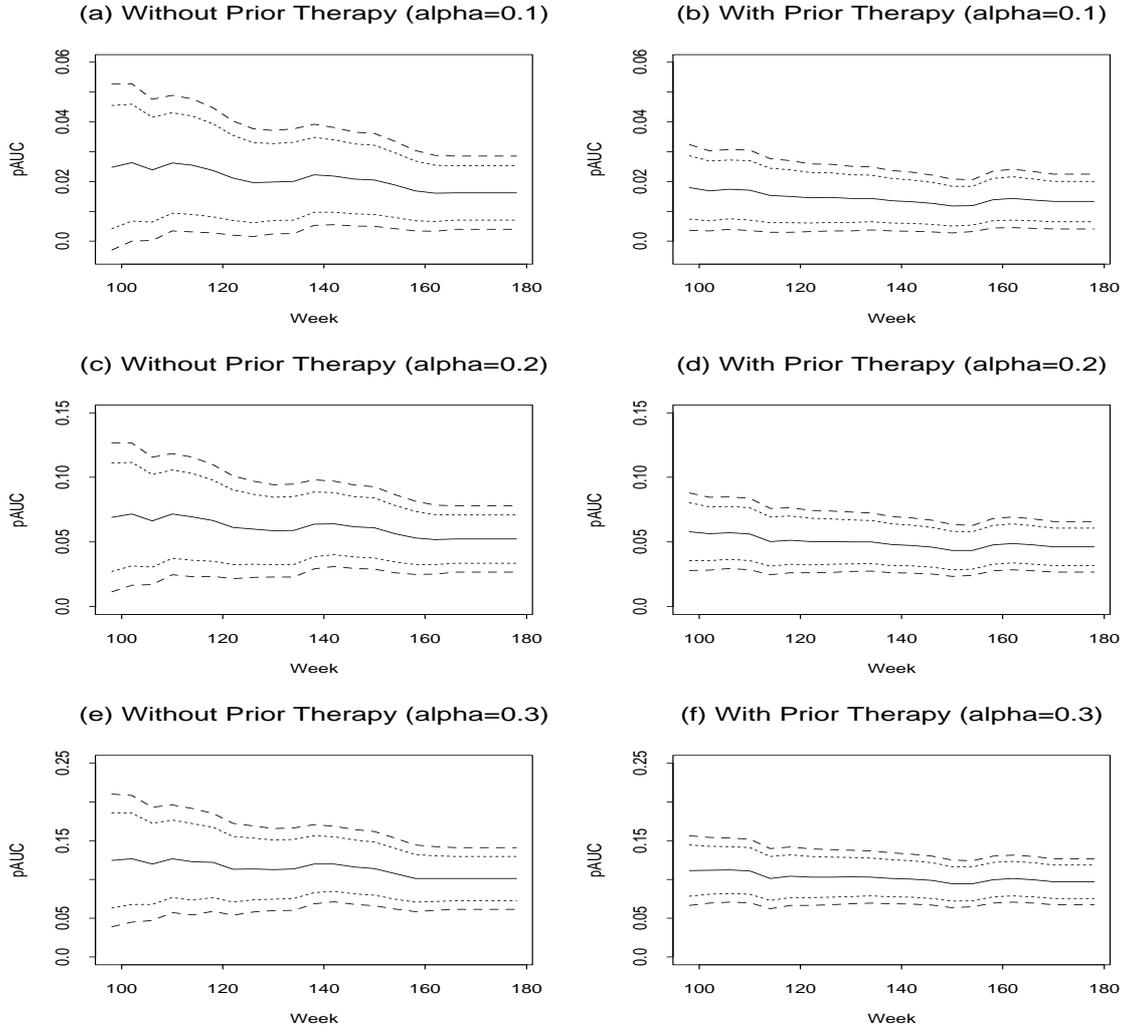}
    \caption{The estimated time-dependent pAUCs (solid curve) and the
    0.95 pointwise confidence intervals (dotted curve) and simultaneous confidence bands (dashed curve).}
\end{figure}

\begin{figure}
\centering
    \includegraphics[width=6in,height=5.5in]{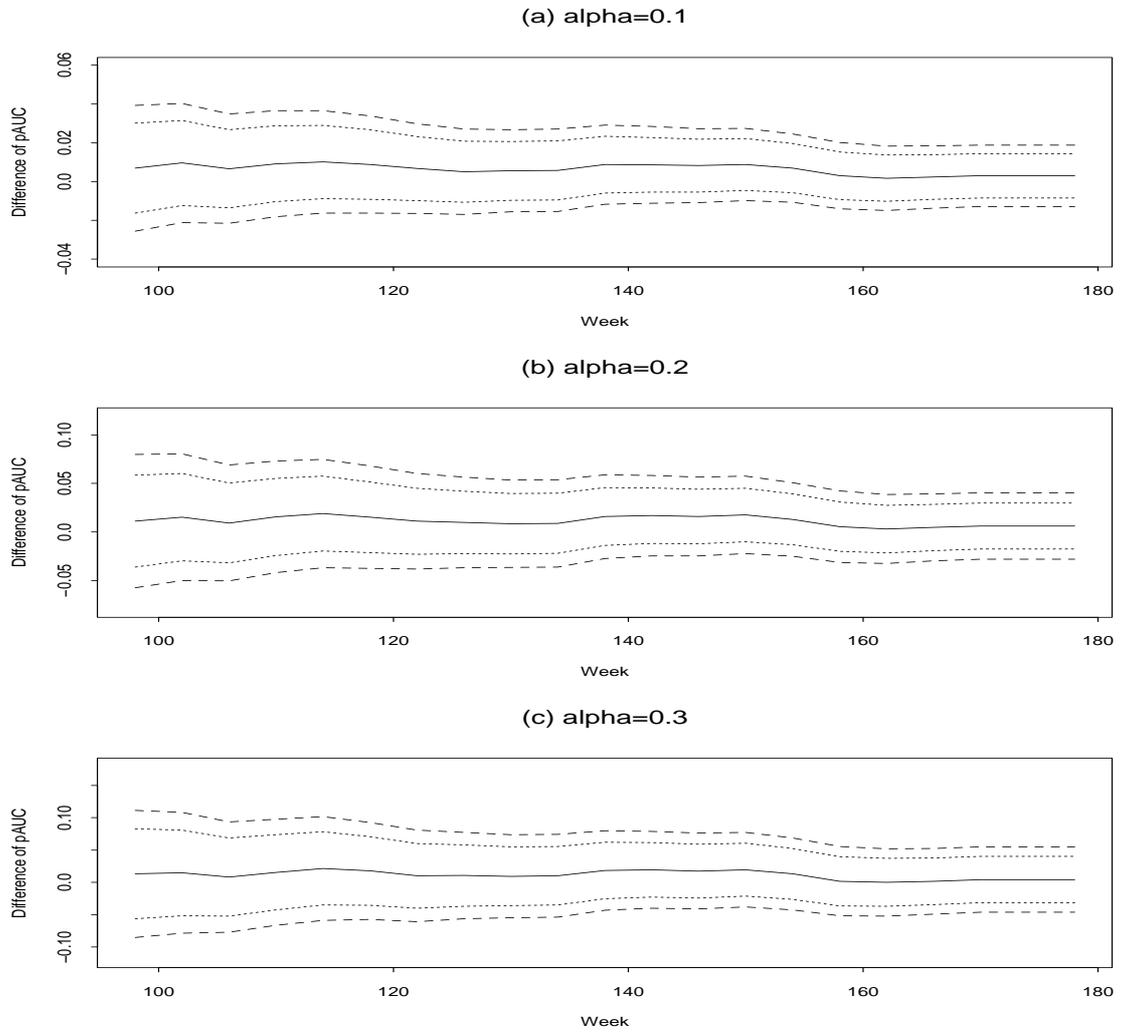}
    \caption{The estimated curves for the difference of the time-dependent pAUCs between non-therapy and therapy patients (solid curve) and the
    0.95 pointwise confidence intervals (dotted curve) and simultaneous confidence bands (dashed curve).}
\end{figure}

\end{document}